\documentclass[prb,twocolumn,showpacs,preprintnumbers,amsmath,amssymb]{revtex4}

\usepackage{graphicx}
\usepackage{dcolumn}
\usepackage{bm}
\usepackage{amsmath}
\usepackage{amssymb}
\usepackage[ansinew]{inputenc}

\newcommand{\ANGS}{\msz \textrm{\AA/s}} 

\newcommand{\GEMN}{$\textrm{Ge}_{0.95}\textrm{Mn}_{0.05}$}
\newcommand{\GRADCM}{^\circ\textrm{C}}
\newcommand{\GRADKM}{\msz \textrm{K}}
\newcommand{\MF}[1]{$\mu_0 \textrm{H} = #1\msz \textrm{T}$}
\newcommand{\msz}{\,} 
\newcommand{\MNGE}{$\textrm{Mn}_5\textrm{Ge}_3$}
\newcommand{\NM}[1]{#1\msz\textrm{nm}}
\newcommand{\TB}{$T_\textrm B$}
\newcommand{\TCMNGE}{$T_\textrm C ^{\textrm{Mn}_5\textrm{Ge}_3}$}

\newcommand{\TS}[1]{$T_\textrm S = #1\GRADCM$}
\newcommand{\TSS}{$T_\textrm S$}

\begin{document}

\title{Magnetic and structural properties of GeMn films: \protect\\ precipitation of intermetallic nanomagnets}

\author{S. Ahlers}
\author{D. Bougeard}
\author{N. Sircar}
\author{G. Abstreiter}
  \affiliation{Walter Schottky Institut\\ Technische Universität München\\ Am Coulombwall 3, D-85748 Garching, Germany}
\author{A. Trampert}
    \affiliation{Paul-Drude-Institut für Festkörperelektronik\\ Hausvogteiplatz 5-7, D-10117 Berlin, Germany}
\author{M. Opel}
\author{R. Gross}
    \affiliation{Walther-Meissner-Institut\\Bayerische Akademie der Wissenschaften\\Walther-Meissner-Str. 8, 85748 Garching, Germany}
\date{\today}

\begin{abstract}
We present a comprehensive study relating the nanostructure of
$\textrm{Ge}_{0.95}\textrm{Mn}_{0.05}$ films  to their magnetic
properties. The formation of ferromagnetic nanometer sized
inclusions in a defect free Ge matrix fabricated by low temperature
molecular beam epitaxy is observed down to substrate temperatures
$T_\textrm S$ as low as $70^\circ\textrm{C}$. A combined
transmission electron microscopy (TEM) and electron energy-loss
spectroscopy (EELS) analysis of the films identifies the inclusions
as precipitates of the ferromagnetic compound
$\textrm{Mn}_5\textrm{Ge}_3$. The volume and amount of these
precipitates decreases with decreasing $T_\textrm S$. Magnetometry of the films containing precipitates reveals distinct temperature ranges: Between the characteristic
ferromagnetic transition temperature of $\textrm{Mn}_5\textrm{Ge}_3$
at approximately room temperature and a lower, $T_\textrm S$
dependent blocking temperature $T_\textrm B$ the magnetic properties
are dominated by superparamagnetism of the
$\textrm{Mn}_5\textrm{Ge}_3$ precipitates. Below $T_\textrm B$, the
magnetic signature of ferromagnetic precipitates with blocked
magnetic moments is observed. At the lowest temperatures, the films
show features characteristic for a metastable state.
\end{abstract}

\maketitle

\section{\label{sec:introduction} Introduction}

Magnetic semiconductors have attracted considerable technological as
well as fundamental scientific interest. Research
has been driven mainly by their possible applicability in
spintronics devices. The GeMn material system is a promising
candidate compatible to the widespread Si semiconductor technology.
Various fabrication techniques have recently been employed to
realise GeMn magnetic semiconductors, including single crystal
growth,\cite{kang:2005PRL} solid phase epitaxy\cite{zeng:2003APL}
and molecular beam epitaxy
(MBE).\cite{park:2001APL,park:2002SCI,li:2005APL,li:2005PRB,d'orazio:2004JMMM,pinto:2005PRB,bihler:2006APL}
Irrespective of the fabrication technique, the formation of
ferromagnetic, intermetallic compounds can occur. Bulk thin
intermetallic films have been observed on
Ge(111),\cite{zeng:2003APL} while phase separation was found on a
$\mu \textrm m$ scale in single crystals\cite{kang:2005PRL} and on a
sub-$\mu \textrm m$ scale in MBE fabricated
samples.\cite{bihler:2006APL} Since the Mn content in ferromagnetic
GeMn intermetallic compounds is of the order of $60\%$ and
above,\cite{predel1996} compound formation will result in large
magnetic moments and a strong influence on the magnetic properties of the system
even if the compounds form on a nanometer scale. An identification of intermetallic compounds in
structural analysis therefore facilitates the interpretation of the magnetic
properties of GeMn films. In particular, structural information is
essential for films with Mn contents of a few percent, where it may
be difficult to unambiguously distinguish a magnetic semiconductor
with intermetallic compounds from a diluted magnetic semiconductor.

In this work we provide new insights into the correlation of the
magnetic and structural properties of \GEMN\ films fabricated with
low temperature MBE. By carefully increasing the substrate
temperature \TSS\ from a temperature as low as $60\GRADCM$, the
onset of the formation of intermetallic compounds was determined. A
combined transmission electron microscopy (TEM) and electron energy-loss spectroscopy (EELS) analysis identifies this compound formation
with the precipitation of nanometer sized \MNGE\ inclusions. The
amount and the size of these inclusions can be tuned with \TSS\
without changing the Mn content of the samples. We therefore present
the first systematic study of the influence of precipitation of
\MNGE\ on the magnetic behaviour of \GEMN\ films. Magnetometry data
reveal a superparamagnetic response of the precipitates between the
ferromagnetic transition temperature \TCMNGE\ of \MNGE\ and a
lower, characteristic temperature \TB, which depends on \TSS\ and thus on the specific nanostructure of the samples.
Below \TB, a blocking process occurs and provides nonzero
magnetisation. At the lowest temperatures,
the system shows the signature of a transition into a metastable state.

\section{\label{sec:experimental}Experimental}
GeMn films were fabricated in a {\sc Riber Siva 32} molecular beam epitaxy machine on intrinsic Ge(100) substrates. The substrates were cleaned \emph{in situ} by annealing over at least 8 hours at a temperature of $400 \GRADCM$, followed by a $30\msz \textrm{min}$ annealing step at $600 \GRADCM$. A $\NM{100}$ intrinsic Ge buffer layer below the GeMn films provided an atomically flat and chemically clean surface. The GeMn films themselves were fabricated at substrate temperatures ranging from \TS{60} to \TS{120} to overcome the low Mn solubility in Ge of approximately $10^{-5}\msz \%$ in thermodynamic equilibrium.\cite{boernstein1984} A Ge rate of $0.08\ANGS$ was chosen to provide good crystalline quality. Mn was introduced by codeposition. All samples presented in this report have a manganese content of nominally $5 \msz \%$ and a GeMn layer thickness of $\NM{200}$. The film quality was monitored \emph{in situ} with reflection high energy electron diffraction (RHEED) and \emph{ex situ} by TEM, x-ray diffraction (XRD) and atomic force microscopy (AFM).

The TEM analysis was carried out on a \textsc{Jeol 3010} microscope operating at 300 kV. Samples were prepared by standard techniques using mechanical grinding, dimpling, and Ar ion-beam milling in a cold stage to avoid sample modifications due to unintentional thermal annealing. Cross-sections were prepared for samples fabricated at $60$, $70$, $100$ and $120\GRADCM$. EELS was used for the chemical analysis of the TEM samples. XRD $\theta-2\theta$ scans were done with a commercial $40\msz\textrm{kV}$ x-ray diffractometer equipped with a $\textrm{Cu K}_{\alpha1}$ cathode.

To investigate the magnetic properties of the films we employed DC and AC superconducting quantum interference device (SQUID) magnetometry with a \textsc{Quantum Design} magnetometer. Several measurement procedures were applied to the samples to account for a possible dependence of the magnetic properties on the measurement history. If not otherwise stated, samples were measured during warmup from 2 to $350 \GRADKM$. Cooling procedures comprise cooling in the maximum available external field of \MF{7} (maximum field cooled, MFC), cooling in the measurement field (field cooled, FC) and cooling without external field (zero field cooled, ZFC). Different measurement fields were used for MFC measurements to extract field induced magnetic properties. Magnetisation loops were taken after maximum field cooldown to the measurement temperature. Both in sample plane and out of sample plane field directions were used.

\section{\label{sec:structural_properties}Structural properties}
RHEED patterns recorded after deposition of the buffer layer reveal perfect ($2\times1$) reconstruction of the Ge surface. To investigate the influence of the low substrate temperatures on the surface reconstruction, we fabricated a reference sample without Mn codeposition at \TS{60}. The intensity of the RHEED patterns recorded after lowering the substrate temperature to $60 \GRADCM$ weakens quickly within less than $\NM{5}$ of low temperature growth. In the following the intensity of the fundamental diffraction spots of the [110] azimuth stays at a constant level, while the [110] half-order diffraction spots disappear. AFM analysis of this reference sample reveals the formation of faceted pits within an otherwise flat surface. In contrary, samples with Mn codeposition exhibit streaky RHEED patterns characteristic for 2D island growth\cite{bratland:2003PRB} for all substrate temperatures, indicating a suppressed pit formation. No pits were visible in AFM images and instead an island covered surface with a root mean square roughness of $\NM{1.5}$ was observed.

\begin{figure}[tbp]
\includegraphics[width=8.6cm]{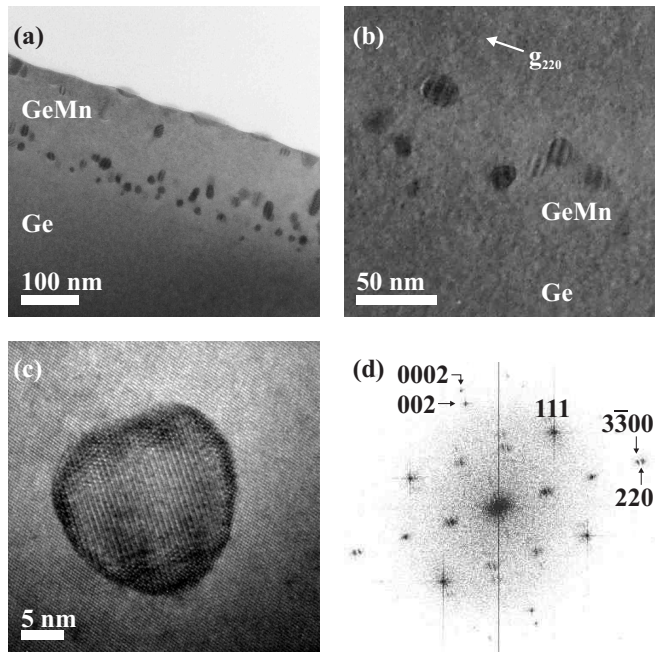}
\caption{\label{fig:TEM_TS120} Cross-sectional TEM micrographs of a sample with \TS{120}. (a) Bright-field cross-sectional TEM overview image. The dark regions correspond to \MNGE\ precipitates. (b) Typical Moiré fringe images of precipitates. (c) High-resolution TEM micrograph of a typical partially coherent precipitate. (d) Fourier transform of (c) showing cubic Ge and hexagonal \MNGE\ lattice reflection indices.}
\end{figure}

Bright-field cross-sectional TEM micrographs show that in spite of substrate temperatures below $120^\circ\textrm{C}$, high quality, dislocation free epitaxy can be achieved. Locally, however,
inclusions are observed for $T_\textrm S \ge 70\GRADCM$ as shown in
Fig.~\ref{fig:TEM_TS120}(a) for the sample with \TS{120}. The
dark regions in bright-field images of these samples correspond to
inclusions in an unperturbed surrounding. For \TS{120}, typical
inclusions are round with an average diameter of \NM{10-20}.

The high-resolution (HR) TEM micrograph in
Fig.~\ref{fig:TEM_TS120}(c) indicates a different crystal structure
and/or orientation of these inclusions compared to the cubic matrix.
The corresponding Fourier transform (FT) in
Fig.~\ref{fig:TEM_TS120}(d) exhibits reflections in addition to the
cubic Ge reflections which can be explained as part of a hexagonal
pattern. The intermetallic phase \MNGE\ is reported to be of
hexagonal $\textrm D8_8$
structure.\cite{castelliz:1953MCCM,forsyth:1990JPCM} Indeed, the FT
pattern in Fig.~\ref{fig:TEM_TS120}(d) is in good agreement with the
one expected for \MNGE\ films. The reflection indices in
Fig.~\ref{fig:TEM_TS120}(d) represent the surrounding, cubic Ge
matrix and in addition the reflexes of the hexagonal inclusions.
Taking the Ge crystal lattice $\textrm a_{Ge} =
5.66\msz\textrm{\AA}$ as a reference, the lattice constants of the
hexagonal inclusions are estimated as $a=5.02\pm 0.05
\msz\textrm{\AA}$ and $c=7.26\pm 0.1 \msz\textrm{\AA}$, which agree
well within the error limits with the strain-free bulk
values\cite{forsyth:1990JPCM} of $a=5.053\msz\textrm{\AA}$ and
$c=7.184\msz\textrm{\AA}$ for \MNGE\ films. Furthermore,
$\theta-2\theta$-XRD scans taken in the Ge $(00\ell)$ direction (not
shown) reveal peaks that correspond to the (002) and (004) reflexes
of bulk \MNGE.\cite{castelliz:1953MCCM,zeng:2003APL} From TEM, XRD
and the magnetometry results presented in Sec.
\ref{sec:magnetic_properties} the inclusions are identified as
precipitation of intermetallic \MNGE. Furthermore, a chemical
analysis performed via EELS with a $10\NM$ spot on the inclusions
results in a very high Mn content for the inclusions, also
indicating the presence of the intermetallic phase. In contrast, the
Mn content in the matrix was found to be less than the detection
limit of $1\%$. No sign of other intermetallic compounds was found
by any of the employed characterisation techniques.

The orientation of the precipitates with respect to the Ge matrix is
observed to be not random but showing the following topotaxial
relation: $(0002)\msz\textrm{\MNGE}\msz\|\msz(002)\msz\textrm{Ge}$ and $\langle11\bar 20$$\rangle\msz\textrm{\MNGE}\msz\|\msz\langle110\rangle\msz\textrm{Ge}$, which agrees with recent results.\cite{bihler:2006APL}
Due to the different crystal structures of the intermetallic
precipitates and the Ge matrix, it is not possible to find parallel lattice planes along all crystal directions. The precipitates are therefore
called partially coherent. It is remarkable that most of the
precipitates appear in the same orientation relation to the matrix
as demonstrated for \TS{120} by the Moiré fringe image in Fig.~\ref{fig:TEM_TS120}(b) where the fringes are almost parallel for all
precipitates. This results in a strong structural anisotropy
regarding the distribution of the hexagonal c-axis of the precipitates compared to the
growth direction.

\begin{figure}[btp]
\includegraphics[width=8.6cm]{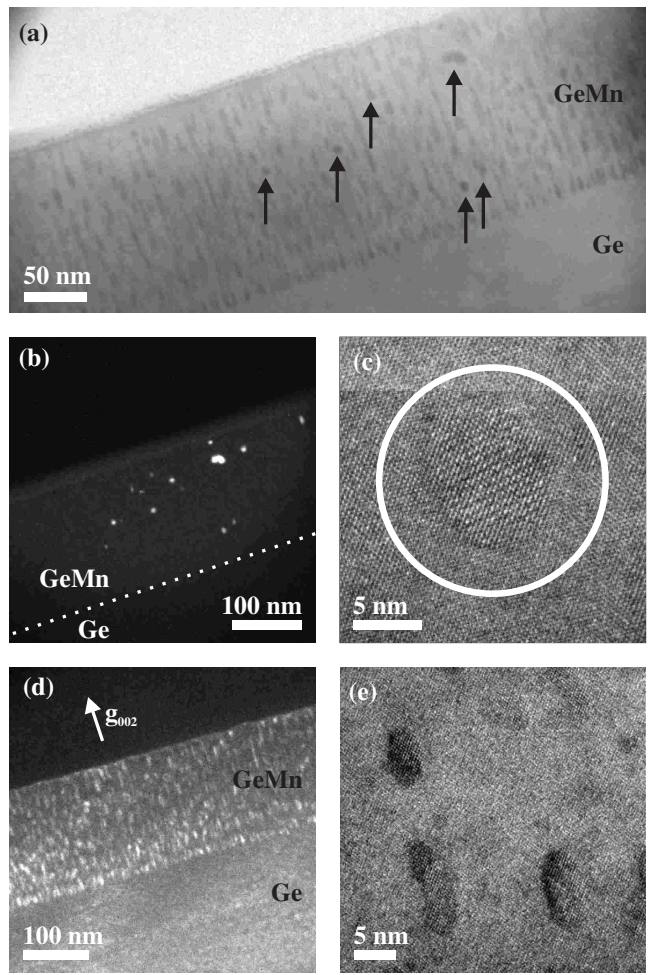}
\caption{\label{fig:TEM_TS70} Cross-sectional TEM micrographs of a sample with \TS{70}. (a) Bright field TEM overview image. Arrows mark \MNGE\ precipitates. (b) Dark-field TEM micrograph recorded after selecting a characteristic reflex of the hexagonal \MNGE\ lattice. The dashed line indicates the interface between GeMn layer and Ge buffer layer. Bright regions correspond to \MNGE\ precipitates. (c) High-resolution TEM of a partially coherent, hexagonal \MNGE\ precipitate, marked by the white circle. (d) Dark-field TEM micrograph recorded after selecting a characteristic reflex of the cubic Ge lattice. Bright regions represent the coherent clusters. (e) High-resolution TEM of typical coherent, cubic clusters. }
\end{figure}

For samples fabricated with $T_\textrm S < 120\GRADCM$ a
different material contrast compared to Fig.~1 (a) is observed in
bright-field TEM micrographs, as illustrated in Fig.~\ref{fig:TEM_TS70}(a) for
\TS{70}. Features resembling the inclusions in Fig.~\ref{fig:TEM_TS120}(a) are marked
with arrows. Their density and diameter is noticeably decreased. In
addition to these features a slight contrast revealing elongated
areas aligned along the growth direction arises. When selecting a
characteristic reflex of the hexagonal \MNGE\ lattice in dark-field
conditions in Fig.~\ref{fig:TEM_TS70}(b), only the inclusions are observed in bright
contrast while the elongated areas disappear. The diameter of the
inclusions is smaller than \NM{9}. A HRTEM image of a typical
inclusion is shown in Fig.~\ref{fig:TEM_TS70}(c). It has a hexagonal
structure and is partially coherent with the matrix as the
precipitates in Fig.~\ref{fig:TEM_TS120}. TEM analysis
performed similarly to the sample with \TS{120} identifies the
inclusions as \MNGE\ precipitates.

The elongated regions aligned along the growth direction can be
visualized in dark-field conditions by selecting a chemically
sensitive reflex of the Ge lattice as shown in
Fig.~\ref{fig:TEM_TS70}(d) for the (002) reflex. They have typical
diameters of $2-5\msz\textrm{nm}$ and lengths of about ten
nanometers. The round shaped \MNGE\ precipitates disappear,
revealing a distinctly different composition and orientation
relation for the elongated, bright areas. The contrast condition
$\textrm g_{002}$ that has to be chosen to observe the bright areas
indicates that they represent regions of higher Mn content compared
to the dark surrounding, the Mn being substitutionally incorporated
on Ge sites. HRTEM micrographs of the sample presented in
Fig.~\ref{fig:TEM_TS70}(e) show dark regions that are similar to the
bright regions both in size and shape. They can be identified with
the bright areas of Fig.~\ref{fig:TEM_TS70}(d). These areas have a
cubic structure and are a coherent part of the host matrix. The dark
and bright areas of Fig.~\ref{fig:TEM_TS70}(d) represent areas of
alternating low and high Mn concentration and thus an inhomogeneous
dispersion of Mn in the Ge matrix. In the following, the bright
areas of Fig.~\ref{fig:TEM_TS70}(d) will be denoted as clusters.
EELS averaging over a $\NM{100}$ spot containing only coherent
clusters results in an average Mn concentration of $\approx 5\%$,
verifying the intended nominal stoichiometry of the
$\textrm{Ge}_{0.95}\textrm{Mn}_{0.05}$ alloys. Since the cluster
sizes are below the EELS resolution limit, no precise information on
the Mn content of the individual clusters can be given.
Nevertheless, an upper limit for the Mn content within the clusters
can be deduced from the dark to bright contrast ratio in
Fig.~\ref{fig:TEM_TS70}(d) and by assuming zero Mn content in the
matrix, leading to a maximum Mn content in the bright areas of
$\approx 15\%$.

TEM analysis shows that only the lowest substrate temperature
sample produced at \TS{60} exhibits no sign for intermetallic
precipitates. It solely contains coherent clusters. The sample with
\TS{70} represents the sample with the lowest substrate temperature
where a marginal amount of hexagonal precipitates can still be found
as illustrated in Fig.~2 (a). TEM analysis on samples with
$60\GRADCM < T_\textrm S \le 120\GRADCM$ reveals an increasing amount
and diameter of these precipitates, while in parallel the fraction
of coherent clusters diminishes. At \TS{120}, only precipitates are
present.

Hence, the variation of $T_\textrm S$ in the epitaxy of
$\textrm{Ge}_{0.95}\textrm{Mn}_{0.05}$ reveals the formation onset
at \TS{70} of the intermetallic compound \MNGE\ as well as the upper limit \TS{120} for the dispersion of Mn in the Ge matrix. Under these epitaxy conditions, this random dispersion is not perfectly homogeneous, leading to the observation of clusters which
are coherently bound to the surrounding and which show enhanced Mn content
compared to the matrix. In contrast, the ferromagnetic \MNGE\ phase appears in
form of unstrained small precipitates. Both precipitates and clusters can
be present in one and the same sample and are thus expected to influence the magnetic properties of the \GEMN\ films,
which are investigated in the following section.

\section{\label{sec:magnetic_properties}Magnetic properties}

\begin{figure}[tp]
\includegraphics[width=8.6cm]{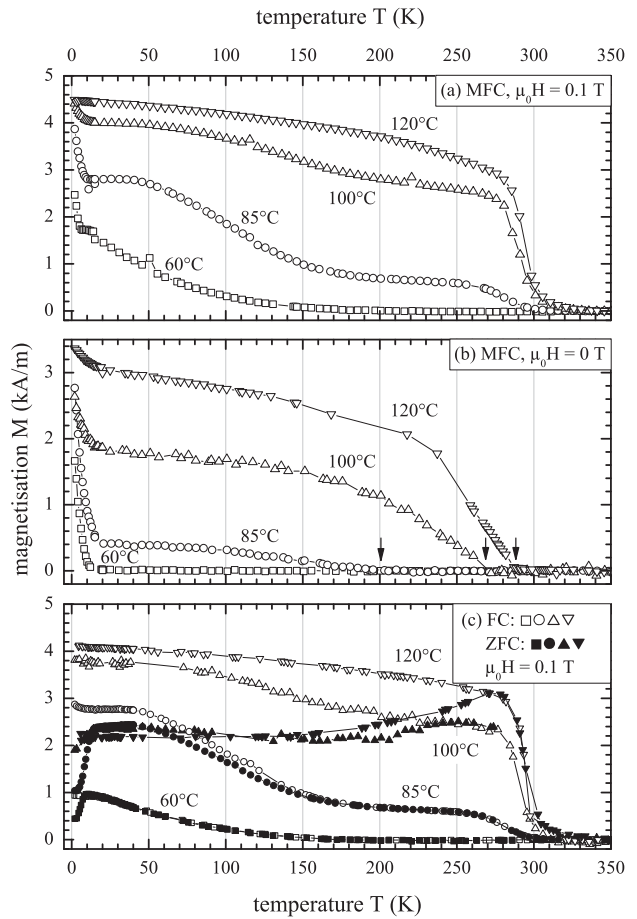}
\caption{\label{fig:SQUID_MT_all} Magnetisation versus temperature curves of samples fabricated with substrate temperatures \TSS\ between $60$ and $120\GRADCM$. External fields were applied in the sample plane. Magnetisation was measured during warmup. (a) MFC, \MF{0.1}. (b) MFC, \MF{0}. The arrows indicate the onset of magnetisation. (c) Field cooled (FC, open symbols) and zero field cooled (ZFC, closed symbols) measurements, \MF{0.1}.}
\end{figure}

Fig.~\ref{fig:SQUID_MT_all} shows the magnetisation of the several \GEMN\ films measured versus
temperature. The curves were measured in different applied magnetic fields after cooling down in zero field (ZFC), the measurement field (FC) or the maximum available field of \MF{7} (MFC). Bulk \MNGE\ is
reported\cite{yamada:1990JPSJ} to be ferromagnetic below
$\textrm{\TCMNGE} = 296\GRADKM$ and to be characterised by a
magnetisation versus temperature curve shape similar to the one observed for the
\TS{120} sample, which, according to Sec.~\ref{sec:structural_properties}, contains only \MNGE\ precipitates.
Therefore the magnetisation onset in Fig.~\ref{fig:SQUID_MT_all}(a)
near room temperature is interpreted as the paramagnetic to
ferromagnetic transition of the individual \MNGE\ precipitates at
\TCMNGE. For all samples containing precipitates, that is samples
with $T_\textrm S > 60 \GRADCM$, a nonzero magnetic signal arises in
the \MF{0.1} MFC measurement of Fig.~\ref{fig:SQUID_MT_all}(a) below
\TCMNGE. Switching off the external field in Fig.~\ref{fig:SQUID_MT_all}(b) reveals that this behaviour is field
induced. The onset of magnetisation now significantly shifts to a
lower, \TSS\ dependent, characteristic temperature. This
temperature, marked by the arrows in Fig.~\ref{fig:SQUID_MT_all}(b),
will hereafter be denoted as \TB.

In contrast to this, the precipitate-free
\TS{60} sample containing only coherent clusters does not show a field induced magnetic signal at
\TCMNGE\ in Fig.~\ref{fig:SQUID_MT_all}(a). In the absence of an
external field, there is no measurable magnetisation in
Fig.~\ref{fig:SQUID_MT_all}(b) above approximately $18\GRADKM$.
Instead, the magnetic signature of the coherent clusters is a field
induced magnetisation up to approximately $200\GRADKM$ as shown in
Fig.~\ref{fig:SQUID_MT_all}(a). Therefore, the contribution of the
coherent clusters on the magnetisation versus temperature curves for $60\GRADCM <
T_\textrm S < 120\GRADCM$ can be experimentally separated from the
contribution of the precipitates by examining the precipitate-free \TS{60} sample.

In FC~/~ZFC measurements in Fig.~\ref{fig:SQUID_MT_all}(c), samples
with $T_\textrm S > 60 \GRADCM$ exhibit a peak in the ZFC curve and
a bifurcation of FC and ZFC curves slightly below the position of the peak in the ZFC magnetisation curve. The
ZFC peak position coincides with \TB. For the \TS{60} sample, no peak or
bifurcation at all is found at high temperatures. The \TS{85} sample
in turn exhibits a bifurcation of FC and ZFC curves, but no peak at \TB\ in
the ZFC curve.

In the measurements of Fig.~\ref{fig:SQUID_MT_all}(a) and (b), a
pronounced, steep increase of magnetisation is visible for samples
with $T_\textrm S < 120\GRADCM$ below approximately $18\GRADKM$. In
the same temperature region, a drop of the ZFC signal compared to a
plateau-like curve form of the FC signal is observed in Fig.~\ref{fig:SQUID_MT_all}(c) for these samples. The steep increase in magnetisation is the only
deviation for the \TS{60} sample from the zero magnetisation in Fig.~\ref{fig:SQUID_MT_all}(b). A thorough analysis of the
precipitate-free \TS{60} sample reveals that this behaviour can be
interpreted as a magnetic signature of the coherent
clusters.\cite{bougeard2006}

\begin{figure}[tbp]
\includegraphics[width=8.6cm]{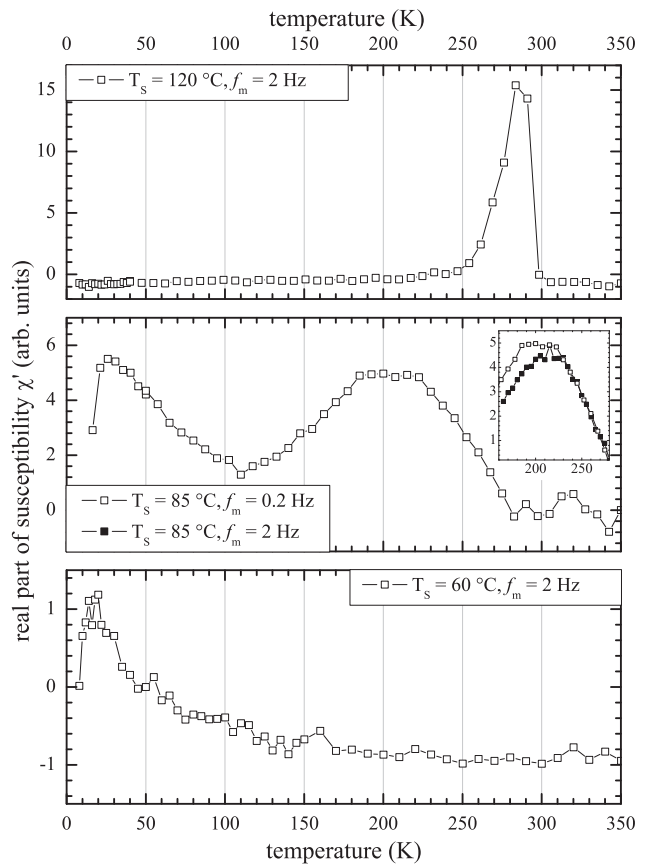}
\caption{\label{fig:SQUID_AC_all} Real part of the AC susceptibility versus temperature. The measurement field $\mu_0 \textrm H = 0.5 \msz\textrm{mT}$ was applied perpendicular to the sample plane. (inset) Shift of the AC susceptibility peak with measurement frequency towards higher temperatures, shown for a sample with \TS{85}.}
\end{figure}

The real part of the AC susceptibility
of the \TS{85} sample presented in Fig.~\ref{fig:SQUID_AC_all}
shows two well separated peaks, one at low temperatures, and one which coincides with the characteristic
temperature \TB. While the first one at low temperatures is also present for the
precipitate-free \TS{60} sample, the latter is not visible in that
sample. Furthermore, the second peak is the only observable peak in the \TS{120} sample which solely contains precipitates, so that we attribute the second peak to the presence of \MNGE\ precipitates, and the first peak to the presence of coherent clusters. The high temperature peak occurs
significantly below \TCMNGE\ for \TS{85} and slightly below \TCMNGE\ for \TS{120}, ruling out a correlation to a \MNGE\
paramagnetic to ferromagnetic transition. The peak position
shifts with the measurement frequency, as shown in the inset of Fig.~\ref{fig:SQUID_AC_all}.

\section{\label{sec:discussion}Discussion}

The magnetic characterisation of the samples shown in Fig.~\ref{fig:SQUID_MT_all} reveals three different temperature regimes. The first regime begins with a field induced onset of magnetisation at \TCMNGE\ in the presence of an external field, as shown in Fig.~\ref{fig:SQUID_MT_all}(a). The second regime is marked by the onset of magnetisation at \TB\ below \TCMNGE\ in the absence of an external field, as indicated by the arrows in Fig.~\ref{fig:SQUID_MT_all}(b). The third regime is found at low temperatures, where a steep increase of magnetisation towards lower temperatures is observed both in Figs. \ref{fig:SQUID_MT_all}(a) and (b). Each of these temperature regions will be discussed separately in the following subsections.

\subsection{Superparamagnetic regime}
\begin{figure}[tbp]
\includegraphics[width=8.6cm]{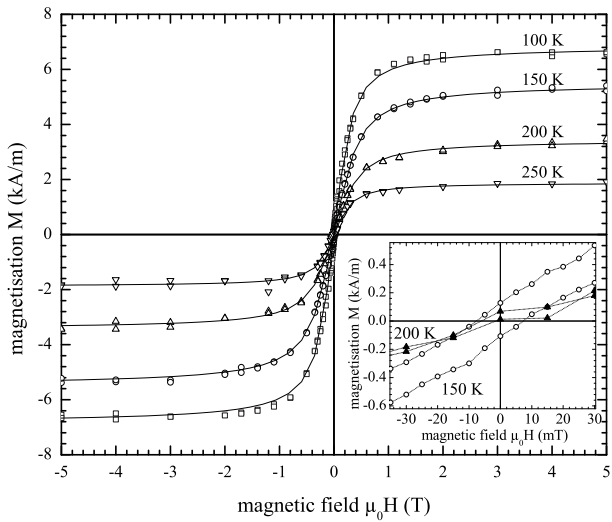}
\caption{\label{fig:SQUID_MH_TS85} Magnetisation curves taken for a \TS{85} sample with external field applied in plane of the sample. Solid lines represent Langevin fits. (inset) Closeup of the same dataset. Hysteresis effects gradually appear below $200\GRADKM$.}
\end{figure}

The highest relevant temperature involved in a complete
description of the magnetic properties of samples with $T_\textrm S > 60\GRADCM$ is the Curie temperature of \MNGE, i.e.
$T_\textrm{C}^\textrm{\MNGE} = 296 \GRADKM$.\cite{yamada:1990JPSJ}
At this temperature, the individual \MNGE\ precipitates turn
ferromagnetic and carry large magnetic moments. The
typical diameter of the precipitates observed in the TEM analysis of Sec.~\ref{sec:structural_properties} is reasonably smaller than the
roughly estimated critical value\cite{kittel:1946PR} of \NM{15} to
display single domain behaviour. Therefore, below \TCMNGE, the precipitates are expected to react freely on an externally applied field like a paramagnet with large magnetic moment, that is like a
superparamagnet. A consequence of the superparamagnetic response of the samples on an applied field is the field induced magnetisation revealed in the magnetisation versus temperature curves of
Fig.~\ref{fig:SQUID_MT_all}. This field induced behaviour
is reversible, so that MFC, FC and ZFC measurements coincide. Reversibility also results in the disappearance of hysteresis in magnetisation loops. These loops can then be described by a
superparamagnetic Langevin function $L(y)$, so that
\begin{eqnarray}
\label{eq:langevin}
M(y) &=& M_\textrm S L(y) \nonumber\\
     &=& M_\textrm{S}\left(\coth{y} - \frac{1}{y}\right) \textrm{ , }y = \frac{\mu \mu_0 H}{k_\textrm{B} T}
\end{eqnarray}
with $M_\textrm{S}$ being the saturation magnetisation and $\mu$ the
magnetic moment of the particles. This is shown in Fig.~\ref{fig:SQUID_MH_TS85} at $T = 200\GRADKM$ and $T = 250\GRADKM$ for the sample with \TS{85}. The Langevin fit at these temperatures results in magnetic moments of the order of $\mu = 2000\mu_\textrm B$. The superparamagnetic regime extends down to \TB, in case of the \TS{85} sample down to $T_\textrm B = 205\GRADKM$. Below \TB, a blocking process discussed in the following subsection occurs.

\subsection{\label{sec:discussion:blocking}Blocking regime}

Bulk \MNGE\ is reported to exhibit uniaxial magnetic anisotropy with respect to
its \textit c-axis.\cite{tawara:1963JPSJ, picozzi:2004PRB} The
presence of such a magnetic anisotropy in an assembly of
superparamagnetic fine particles leads to the existence of an
anisotropy energy barrier $E_\textrm A$,
\begin{equation}
E_\textrm A = K V \sin^2(\theta)
\end{equation}
with $K$ being the anisotropy energy density, $V$ the particle
volume and $\theta$ the angle between magnetic moment and anisotropy
axis.\cite{jacobs1963} The energy barrier does not affect the system
at the comparably high temperatures of the superparamagnetic regime, where the anisotropy energy barrier is smaller than the thermal energy and hence is easily overcome by thermal activation.
However, if the temperature is lowered below a characteristic
blocking temperature \TB, the energy barrier blocks the magnetic
moments in a direction either parallel or antiparallel to the
anisotropy axis.

According to the theory of Néel and Brown, the direction of magnetisation of superparamagnetic particles undergoes a sort of Brownian motion.\cite{jacobs1963} In the presence of an energy barrier, these thermal fluctuations take place with a relaxation rate that is given by
\begin{equation}
\label{eq:sp_relaxation_rate}
f = f_0\exp{\left(-\frac{KV}{k_\textrm B T}\right).}
\end{equation}
$f_0$ is a typical attempt frequency of the order of $10^9\msz\textrm{Hz}$.\cite{brown:1963PR,brown:1959JAP} With decreasing temperature the fluctuations slow down until the magnetic moments eventually become stable below the temperature \TB\ on the characteristic time scale $1/f$ of an experiment. The associated temperature is denoted as the blocking temperature \TB. An AC susceptibility experiment with a measurement frequency $f_\textrm{m}$ directly probes this blocking process. A peak in the measured susceptibility versus temperature curve is expected for the temperature \TB\ where the measurement frequency $f_\textrm{m}$ is equal to the fluctuation frequency $f$. According to this model different measurement frequencies are associated with different values of \TB. Hence, a shift of the peak position in the susceptibility versus temperature curves is obtained as shown in the inset of Fig.~\ref{fig:SQUID_AC_all} for a sample with \TS{85}. The measured value of the relative shift per frequency decade,
\begin{equation}
\frac{\Delta T_\textrm B}{T_\textrm B \Delta \log f}
\end{equation}
equals $0.075$ and is close to the value of $0.1$ typically observed in superparamagnetic systems. Following Bean and Livingston,\cite{bean:1959JAP} the thermal fluctuations can be considered as stable if the relaxation time of the fluctuations $\tau = 1 / f$ is of the order of $10^2\msz\textrm s$, that is, if
\begin{equation}
\label{eq:anisotropy_constant}
25k_\textrm B T_\textrm B = KV\textrm{.}
\end{equation}
Employing the particle dimensions found in the TEM analysis of Sec.~\ref{sec:structural_properties} and assuming a spherical particle shape in Eq. (\ref{eq:anisotropy_constant}), we can estimate values for the anisotropy constant $K$ of the precipitates from the blocking temperatures observed in Fig.~\ref{fig:SQUID_MT_all}(b). The results of this analysis are summarised in Tab.~\ref{tab:blocking_temperatures}.

\begin{table}
\caption{\label{tab:blocking_temperatures}Blocking temperatures for
samples with precipitates extracted from the measurements of Fig.~\ref{fig:SQUID_MT_all}, mean precipitate diameter and estimated anisotropy constant $K$.}
\begin{ruledtabular}
\begin{tabular}{lrrrr}
\TSS&$70\GRADCM$&$85\GRADCM$&$100\GRADCM$&$120\GRADCM$\\
\hline
\TB\ in K (from Fig.~\ref{fig:SQUID_MT_all}(b)) & 118\footnote{not shown in Fig.~\ref{fig:SQUID_MT_all}}  & 205 & 269 & 287\\
\TB\ in K (from Fig.~\ref{fig:SQUID_MT_all}(c)) &-& 198 & 258 & 278\\
mean diameter in nm (TEM) &9.0& - & 11.4 & 13.6\\
$K$\footnote{obtained from Eq. (\ref{eq:anisotropy_constant}) with \TB\ values of Fig.~\ref{fig:SQUID_MT_all}(b)} in $10^6\msz\textrm{erg}/ \textrm{cm}^3$ &1.1& - & 1.2 & 0.8\\
\end{tabular}
\end{ruledtabular}
\end{table}

In the \MF{0} MFC measurements in Fig.~\ref{fig:SQUID_MT_all}(b), the precipitate moments are aligned by
the maximum field cooldown. Below \TB, they are blocked by the anisotropy energy
barrier on the typical time scale of the experiment, so that their magnetic behaviour is no longer reversible and an overall nonzero magnetisation is observable. As shown in Tab.~\ref{tab:blocking_temperatures}, \TB\ increases with increasing \TSS\ and thus with increasing precipitate diameter. The blocking process itself does
not provide a spontaneous overall magnetisation, as parallel or
antiparallel alignment of the magnetic moments with respect to the
anisotropy axis is equally probable.\cite{hesse:2000JMMM} This is
observed when magnetisation is measured during zero field cooldown
(not shown), where no overall nonzero magnetisation is detected. The absence of reversibility below \TB\ results in a bifurcation of FC and ZFC curves below \TB, as
well as in a peak at \TB\ in ZFC measurements in Fig.~\ref{fig:SQUID_MT_all}(c). Hysteresis effects gradually arising below \TB\ are a further consequence of the blocking of the magnetic supermoments. This is shown in the inset of Fig.~\ref{fig:SQUID_MH_TS85} for a \TS{85} sample with $T_\textrm B = 205\GRADKM$.

For each substrate temperature the TEM analysis reveals a certain precipitate size
distribution. Due to the correlation of precipitate
volume and blocking temperature, this results in a broad temperature range of
the blocking process and thus in a broad peak around an average \TB\ in the AC susceptibility measurements. For the sample with \TS{85}, this peak expands between approximately $100$ and $275\GRADKM$ as shown in Fig.~\ref{fig:SQUID_AC_all}. The broad temperature range of the blocking process leads to a situation where, depending on the
individual particle volume, both blocked and unblocked moments are
present at a given temperature close to the experimentally observed average \TB. This explains why physically reasonable Langevin fits of the reversible part of the magnetisation loops in Fig.~\ref{fig:SQUID_MH_TS85} can be performed below $T_\textrm B = 205\GRADKM$ down to $100\GRADKM$. The deduced magnetic moments range from $1200\msz\mu_\textrm B$ at $100\GRADKM$ to $2900\msz\mu_\textrm B$ at $250\GRADKM$ and increase with measurement temperature. Larger particles are
blocked up to higher temperatures, so that for lower temperatures
only the smaller particles are probed by the Langevin function and
smaller magnetic moments are found. To observe temperature
independent magnetic supermoments in magnetisation loops,
measurements would have to be performed at temperatures well above
the blocking temperature, which is hindered by \TCMNGE\ naturally
setting a limit to the superparamagnetic regime.

In samples that contain both clusters and precipitates, that is for
$60\GRADCM < T_\textrm S < 120\GRADCM$, the magnetic signal of the
coherent clusters is overlaid over the magnetic signature of
the intermetallic precipitates. In an externally applied field as in
Fig.~\ref{fig:SQUID_MT_all}(a), this results in magnetisation versus temperature curves that can be regarded as a superposition of those obtained for the \TS{60} sample and the ones obtained for higher \TSS\ samples. Also the temperature of the magnetisation onset of the \TS{60} sample is visible in the higher \TSS\ samples as a
shoulder in the magnetisation versus temperature curves. The height of this shoulder decreases
with \TSS, since the coherent cluster content decreases with higher
substrate temperatures, as discussed in Sec.~\ref{sec:structural_properties}. Without an external field as in
Fig.~\ref{fig:SQUID_MT_all}(b), only the signal coming from the
\MNGE\ precipitates is observable above $18\GRADKM$, since the clusters do not exhibit
an overall nonzero magnetic signal in zero field conditions.
The \TS{85} sample contains considerable amounts of both coherent
clusters and \MNGE\ precipitates. The peak at \TB\ in the ZFC measurements of Fig.~3c is suppressed for this sample, while it is pronounced for samples with higher \TSS\ and therefore lower cluster content. We interpret this suppression to result from the ZFC curve for \TS{120}, characteristic
for the contribution of the precipitates, being overlaid by the ZFC curve
for \TS{60}, characteristic for the clusters.

\subsection{Metastable regime}
For temperatures below $18\GRADKM$, the precipitate-free \TS{60} sample exhibits magnetic signatures of a transition into a metastable state, which we proposed\cite{bougeard2006} to be a consequence of the formation of coherent clusters. These magnetic signatures are observed for samples with $T_\textrm S > 60\GRADCM$ in the same temperature range. They appear as a drop
of the ZFC signal at low temperatures compared to a plateau-like
curve form of the FC signal of Fig.~\ref{fig:SQUID_MT_all}(c), and as a steep increase in the MFC measurements of Fig.~\ref{fig:SQUID_MT_all}(a) and (b). Furthermore, magnetisation relaxation in time dependent measurements (not shown) and a peak in the AC susceptibility, as shown in Fig.~\ref{fig:SQUID_AC_all}, occur at this temperature range. This behaviour is visible for all samples containing coherent clusters and gets more and more suppressed as \TSS\ is increased, that is as the amount of coherent clusters in the films is reduced. From this trend we
conclude that the features characteristic of a metastable state
observed for samples with $T_\textrm S > 60\GRADCM$ are due to the
presence of the inhomogeneous dispersion of Mn in the Ge matrix.

\section{Conclusion}
In conclusion, we have presented a study of the magnetic and
structural properties of $\textrm{Ge}_{0.95}\textrm{Mn}_{0.05}$
films fabricated by low temperature MBE at substrate temperatures
\TSS\ ranging from $60\GRADCM$ to $120\GRADCM$. An extensive TEM analysis reveals the precipitation of nanometer sized
intermetallic \MNGE\ compounds for substrate temperatures $T_\textrm
S \ge 70\GRADCM$. The amount and size of these inclusions can be
tuned with the substrate temperature without changing the Mn
content of the samples. Precipitation can be suppressed at \TS{60}. Therefore, we were able to systematically
investigate and identify the influence of precipitation on the magnetic behaviour
of the films. The magnetic signature of the precipitates is a field induced, superparamagnetic signal between the ferromagnetic transition temperature $\textrm{\TCMNGE}=296\GRADKM$ of \MNGE\ and a lower, characteristic temperature \TB. Below \TB, nonzero magnetisation in the absence of an external field, as well as hysteresis in magnetisation loops is observed, which is due to a blocking process of the superparamagnetic precipitates. Therefore our study indicates that reports on hysteresis in magnetisation loops\cite{d'orazio:2004JMMM,pinto:2005PRB} and a field induced magnetisation onset near room temperature\cite{park:2002SCI} in possibly diluted magnetic semiconductors might be a result of the presence of precipitates in the films. At low temperatures, samples with $70 \GRADCM \le T_\textrm S < 120\GRADCM$ undergo a transition into a metastable state that is interpreted as a signature of an inhomogeneous Mn dispersion in the Ge matrix observed in addition to the precipitates.

\begin{acknowledgments}
This work was funded by Deutsche Forschungsgemeinschaft via SFB 631. The author gratefully acknowledge stimulating discussions with C. Bihler, C. Jäger and M. S. Brandt and EELS measurements by X. Kong.
\end{acknowledgments}


\begin{thebibliography}{24}
\expandafter\ifx\csname natexlab\endcsname\relax\def\natexlab#1{#1}\fi
\expandafter\ifx\csname bibnamefont\endcsname\relax
  \def\bibnamefont#1{#1}\fi
\expandafter\ifx\csname bibfnamefont\endcsname\relax
  \def\bibfnamefont#1{#1}\fi
\expandafter\ifx\csname citenamefont\endcsname\relax
  \def\citenamefont#1{#1}\fi
\expandafter\ifx\csname url\endcsname\relax
  \def\url#1{\texttt{#1}}\fi
\expandafter\ifx\csname urlprefix\endcsname\relax\def\urlprefix{URL }\fi
\providecommand{\bibinfo}[2]{#2}
\providecommand{\eprint}[2][]{\url{#2}}

\bibitem[{\citenamefont{Kang et~al.}(2005)\citenamefont{Kang, Kim, Wi, Lee,
  Choi, Cho, Han, Kim, Song, Shin et~al.}}]{kang:2005PRL}
\bibinfo{author}{\bibfnamefont{J.-S.} \bibnamefont{Kang}},
  \bibinfo{author}{\bibfnamefont{G.}~\bibnamefont{Kim}},
  \bibinfo{author}{\bibfnamefont{S.~C.} \bibnamefont{Wi}},
  \bibinfo{author}{\bibfnamefont{S.~S.} \bibnamefont{Lee}},
  \bibinfo{author}{\bibfnamefont{S.}~\bibnamefont{Choi}},
  \bibinfo{author}{\bibfnamefont{S.}~\bibnamefont{Cho}},
  \bibinfo{author}{\bibfnamefont{S.~W.} \bibnamefont{Han}},
  \bibinfo{author}{\bibfnamefont{K.~H.} \bibnamefont{Kim}},
  \bibinfo{author}{\bibfnamefont{H.~J.} \bibnamefont{Song}},
  \bibinfo{author}{\bibfnamefont{H.~J.} \bibnamefont{Shin}},
  \bibnamefont{et~al.}, \bibinfo{journal}{Phys. Rev. Lett.}
  \textbf{\bibinfo{volume}{94}}, \bibinfo{pages}{147202}
  (\bibinfo{year}{2005}).

\bibitem[{\citenamefont{Zeng et~al.}(2003)\citenamefont{Zeng, Erwin, Feldman,
  Li, Jin, Song, Thompson, and Weitering}}]{zeng:2003APL}
\bibinfo{author}{\bibfnamefont{C.~G.} \bibnamefont{Zeng}},
  \bibinfo{author}{\bibfnamefont{S.~C.} \bibnamefont{Erwin}},
  \bibinfo{author}{\bibfnamefont{L.~C.} \bibnamefont{Feldman}},
  \bibinfo{author}{\bibfnamefont{A.~P.} \bibnamefont{Li}},
  \bibinfo{author}{\bibfnamefont{R.}~\bibnamefont{Jin}},
  \bibinfo{author}{\bibfnamefont{Y.}~\bibnamefont{Song}},
  \bibinfo{author}{\bibfnamefont{J.~R.} \bibnamefont{Thompson}},
  \bibnamefont{and} \bibinfo{author}{\bibfnamefont{H.~H.}
  \bibnamefont{Weitering}}, \bibinfo{journal}{Appl. Phys. Lett.}
  \textbf{\bibinfo{volume}{83}}, \bibinfo{pages}{5002} (\bibinfo{year}{2003}).

\bibitem[{\citenamefont{Park et~al.}(2001)\citenamefont{Park, Wilson, Hanbicki,
  Mattson, Ambrose, and Spanos}}]{park:2001APL}
\bibinfo{author}{\bibfnamefont{Y.~D.} \bibnamefont{Park}},
  \bibinfo{author}{\bibfnamefont{A.}~\bibnamefont{Wilson}},
  \bibinfo{author}{\bibfnamefont{A.~T.} \bibnamefont{Hanbicki}},
  \bibinfo{author}{\bibfnamefont{J.~E.} \bibnamefont{Mattson}},
  \bibinfo{author}{\bibfnamefont{T.}~\bibnamefont{Ambrose}}, \bibnamefont{and}
  \bibinfo{author}{\bibfnamefont{G.}~\bibnamefont{Spanos}},
  \bibinfo{journal}{Appl. Phys. Lett.} \textbf{\bibinfo{volume}{78}},
  \bibinfo{pages}{2739} (\bibinfo{year}{2001}).

\bibitem[{\citenamefont{Park et~al.}(2002)\citenamefont{Park, Hanbicki, Erwin,
  Hellberg, Sullivan, Mattson, Ambrose, Wilson, Spanos, and
  Jonker}}]{park:2002SCI}
\bibinfo{author}{\bibfnamefont{Y.~D.} \bibnamefont{Park}},
  \bibinfo{author}{\bibfnamefont{A.~T.} \bibnamefont{Hanbicki}},
  \bibinfo{author}{\bibfnamefont{S.~C.} \bibnamefont{Erwin}},
  \bibinfo{author}{\bibfnamefont{C.~S.} \bibnamefont{Hellberg}},
  \bibinfo{author}{\bibfnamefont{J.~M.} \bibnamefont{Sullivan}},
  \bibinfo{author}{\bibfnamefont{J.~E.} \bibnamefont{Mattson}},
  \bibinfo{author}{\bibfnamefont{T.~F.} \bibnamefont{Ambrose}},
  \bibinfo{author}{\bibfnamefont{A.}~\bibnamefont{Wilson}},
  \bibinfo{author}{\bibfnamefont{G.}~\bibnamefont{Spanos}}, \bibnamefont{and}
  \bibinfo{author}{\bibfnamefont{B.~T.} \bibnamefont{Jonker}},
  \bibinfo{journal}{Science} \textbf{\bibinfo{volume}{295}},
  \bibinfo{pages}{651} (\bibinfo{year}{2002}).

\bibitem[{\citenamefont{Li et~al.}(2005{\natexlab{a}})\citenamefont{Li, Shen,
  Thompson, and Weitering}}]{li:2005APL}
\bibinfo{author}{\bibfnamefont{A.~P.} \bibnamefont{Li}},
  \bibinfo{author}{\bibfnamefont{J.}~\bibnamefont{Shen}},
  \bibinfo{author}{\bibfnamefont{J.~R.} \bibnamefont{Thompson}},
  \bibnamefont{and} \bibinfo{author}{\bibfnamefont{H.~H.}
  \bibnamefont{Weitering}}, \bibinfo{journal}{Appl. Phys. Lett.}
  \textbf{\bibinfo{volume}{86}}, \bibinfo{pages}{152507}
  (\bibinfo{year}{2005}{\natexlab{a}}).

\bibitem[{\citenamefont{Li et~al.}(2005{\natexlab{b}})\citenamefont{Li,
  Wendelken, Shen, Feldman, Thompson, and Weitering}}]{li:2005PRB}
\bibinfo{author}{\bibfnamefont{A.~P.} \bibnamefont{Li}},
  \bibinfo{author}{\bibfnamefont{J.~F.} \bibnamefont{Wendelken}},
  \bibinfo{author}{\bibfnamefont{J.}~\bibnamefont{Shen}},
  \bibinfo{author}{\bibfnamefont{L.~C.} \bibnamefont{Feldman}},
  \bibinfo{author}{\bibfnamefont{J.~R.} \bibnamefont{Thompson}},
  \bibnamefont{and} \bibinfo{author}{\bibfnamefont{H.~H.}
  \bibnamefont{Weitering}}, \bibinfo{journal}{Phys. Rev. B}
  \textbf{\bibinfo{volume}{72}}, \bibinfo{pages}{195205}
  (\bibinfo{year}{2005}{\natexlab{b}}).

\bibitem[{\citenamefont{D'Orazio et~al.}(2004)\citenamefont{D'Orazio, Lucari,
  Pinto, Morresi, and Murri}}]{d'orazio:2004JMMM}
\bibinfo{author}{\bibfnamefont{F.}~\bibnamefont{D'Orazio}},
  \bibinfo{author}{\bibfnamefont{F.}~\bibnamefont{Lucari}},
  \bibinfo{author}{\bibfnamefont{N.}~\bibnamefont{Pinto}},
  \bibinfo{author}{\bibfnamefont{L.}~\bibnamefont{Morresi}}, \bibnamefont{and}
  \bibinfo{author}{\bibfnamefont{R.}~\bibnamefont{Murri}}, \bibinfo{journal}{J.
  Magn. Magn. Mater.} \textbf{\bibinfo{volume}{272-276}}, \bibinfo{pages}{2006}
  (\bibinfo{year}{2004}).

\bibitem[{\citenamefont{Pinto et~al.}(2005)\citenamefont{Pinto, Morresi,
  Ficcadenti, Murri, D'Orazio, Lucari, Boarino, and Amato}}]{pinto:2005PRB}
\bibinfo{author}{\bibfnamefont{N.}~\bibnamefont{Pinto}},
  \bibinfo{author}{\bibfnamefont{L.}~\bibnamefont{Morresi}},
  \bibinfo{author}{\bibfnamefont{M.}~\bibnamefont{Ficcadenti}},
  \bibinfo{author}{\bibfnamefont{R.}~\bibnamefont{Murri}},
  \bibinfo{author}{\bibfnamefont{F.}~\bibnamefont{D'Orazio}},
  \bibinfo{author}{\bibfnamefont{F.}~\bibnamefont{Lucari}},
  \bibinfo{author}{\bibfnamefont{L.}~\bibnamefont{Boarino}}, \bibnamefont{and}
  \bibinfo{author}{\bibfnamefont{G.}~\bibnamefont{Amato}},
  \bibinfo{journal}{Phys. Rev. B} \textbf{\bibinfo{volume}{72}},
  \bibinfo{pages}{165203} (\bibinfo{year}{2005}).

\bibitem[{\citenamefont{Bihler et~al.}(2006)\citenamefont{Bihler, Jaeger,
  Vallaitis, Gjukic, Brandt, Pippel, Woltersdorf, and Gösele}}]{bihler:2006APL}
\bibinfo{author}{\bibfnamefont{C.}~\bibnamefont{Bihler}},
  \bibinfo{author}{\bibfnamefont{C.}~\bibnamefont{Jaeger}},
  \bibinfo{author}{\bibfnamefont{T.}~\bibnamefont{Vallaitis}},
  \bibinfo{author}{\bibfnamefont{M.}~\bibnamefont{Gjukic}},
  \bibinfo{author}{\bibfnamefont{M.~S.} \bibnamefont{Brandt}},
  \bibinfo{author}{\bibfnamefont{E.}~\bibnamefont{Pippel}},
  \bibinfo{author}{\bibfnamefont{J.}~\bibnamefont{Woltersdorf}},
  \bibnamefont{and} \bibinfo{author}{\bibfnamefont{U.}~\bibnamefont{Gösele}},
  \bibinfo{journal}{Appl. Phys. Lett.} \textbf{\bibinfo{volume}{88}},
  \bibinfo{pages}{112506} (\bibinfo{year}{2006}).

\bibitem[{\citenamefont{Predel}(1996)}]{predel1996}
\bibinfo{author}{\bibfnamefont{B.}~\bibnamefont{Predel}}, in
  \emph{\bibinfo{booktitle}{Landolt-Börnstein - Group IV: Physical Chemistry}}
  (\bibinfo{publisher}{Springer}, \bibinfo{address}{Berlin, Heidelberg},
  \bibinfo{year}{1996}), vol.~\bibinfo{volume}{5}.

\bibitem[{\citenamefont{Mühlbauer}(1984)}]{boernstein1984}
\bibinfo{author}{\bibfnamefont{A.}~\bibnamefont{Mühlbauer}}, in
  \emph{\bibinfo{booktitle}{{L}andolt-{B}örnstein - Group III: Crystal and
  Solid State Physics}} (\bibinfo{publisher}{Springer},
  \bibinfo{address}{Berlin, Heidelberg, New York, Tokio},
  \bibinfo{year}{1984}), vol. \bibinfo{volume}{17/C}.

\bibitem[{\citenamefont{Bratland et~al.}(2003)\citenamefont{Bratland, Foo,
  Soares, Spila, Desjardins, and Greene}}]{bratland:2003PRB}
\bibinfo{author}{\bibfnamefont{K.~A.} \bibnamefont{Bratland}},
  \bibinfo{author}{\bibfnamefont{Y.~L.} \bibnamefont{Foo}},
  \bibinfo{author}{\bibfnamefont{J.~A. N.~T.} \bibnamefont{Soares}},
  \bibinfo{author}{\bibfnamefont{T.}~\bibnamefont{Spila}},
  \bibinfo{author}{\bibfnamefont{P.}~\bibnamefont{Desjardins}},
  \bibnamefont{and} \bibinfo{author}{\bibfnamefont{J.~E.}
  \bibnamefont{Greene}}, \bibinfo{journal}{Phys. Rev. B}
  \textbf{\bibinfo{volume}{67}}, \bibinfo{pages}{125322}
  (\bibinfo{year}{2003}).

\bibitem[{\citenamefont{Castelliz}(1953)}]{castelliz:1953MCCM}
\bibinfo{author}{\bibfnamefont{L.}~\bibnamefont{Castelliz}},
  \bibinfo{journal}{Monatsh. Chem. / Chem. Monthly}
  \textbf{\bibinfo{volume}{84}}, \bibinfo{pages}{765} (\bibinfo{year}{1953}).

\bibitem[{\citenamefont{Forsyth and Brown}(1990)}]{forsyth:1990JPCM}
\bibinfo{author}{\bibfnamefont{J.~B.} \bibnamefont{Forsyth}} \bibnamefont{and}
  \bibinfo{author}{\bibfnamefont{P.~J.} \bibnamefont{Brown}},
  \bibinfo{journal}{J. Phys.: Condens. Matter} \textbf{\bibinfo{volume}{2}},
  \bibinfo{pages}{2713} (\bibinfo{year}{1990}).

\bibitem[{\citenamefont{Yamada}(1990)}]{yamada:1990JPSJ}
\bibinfo{author}{\bibfnamefont{N.}~\bibnamefont{Yamada}}, \bibinfo{journal}{J.
  Phys. Soc. Jpn.} \textbf{\bibinfo{volume}{59}}, \bibinfo{pages}{273}
  (\bibinfo{year}{1990}).

\bibitem[{\citenamefont{Bougeard et~al.}(2006)\citenamefont{Bougeard, Ahlers,
  Trampert, Sircar, and Abstreiter}}]{bougeard2006}
\bibinfo{author}{\bibfnamefont{D.}~\bibnamefont{Bougeard}},
  \bibinfo{author}{\bibfnamefont{S.}~\bibnamefont{Ahlers}},
  \bibinfo{author}{\bibfnamefont{A.}~\bibnamefont{Trampert}},
  \bibinfo{author}{\bibfnamefont{N.}~\bibnamefont{Sircar}}, \bibnamefont{and}
  \bibinfo{author}{\bibfnamefont{G.}~\bibnamefont{Abstreiter}}
  (\bibinfo{year}{2006}), \bibinfo{note}{unpublished}.

\bibitem[{\citenamefont{Kittel}(1946)}]{kittel:1946PR}
\bibinfo{author}{\bibfnamefont{C.}~\bibnamefont{Kittel}},
  \bibinfo{journal}{Phys. Rev.} \textbf{\bibinfo{volume}{70}},
  \bibinfo{pages}{965} (\bibinfo{year}{1946}).

\bibitem[{\citenamefont{Tawara and Sato}(1963)}]{tawara:1963JPSJ}
\bibinfo{author}{\bibfnamefont{Y.}~\bibnamefont{Tawara}} \bibnamefont{and}
  \bibinfo{author}{\bibfnamefont{K.}~\bibnamefont{Sato}}, \bibinfo{journal}{J.
  Phys. Soc. Jpn.} \textbf{\bibinfo{volume}{18}}, \bibinfo{pages}{773}
  (\bibinfo{year}{1963}).

\bibitem[{\citenamefont{Picozzi et~al.}(2004)\citenamefont{Picozzi, Continenza,
  and Freeman}}]{picozzi:2004PRB}
\bibinfo{author}{\bibfnamefont{S.}~\bibnamefont{Picozzi}},
  \bibinfo{author}{\bibfnamefont{A.}~\bibnamefont{Continenza}},
  \bibnamefont{and} \bibinfo{author}{\bibfnamefont{A.~J.}
  \bibnamefont{Freeman}}, \bibinfo{journal}{Phys. Rev. B}
  \textbf{\bibinfo{volume}{70}}, \bibinfo{eid}{235205} (\bibinfo{year}{2004}).

\bibitem[{\citenamefont{Jacobs and Bean}(1963)}]{jacobs1963}
\bibinfo{author}{\bibfnamefont{I.~S.} \bibnamefont{Jacobs}} \bibnamefont{and}
  \bibinfo{author}{\bibfnamefont{C.~P.} \bibnamefont{Bean}}, in
  \emph{\bibinfo{booktitle}{Magnetism}}, edited by
  \bibinfo{editor}{\bibfnamefont{G.~T.} \bibnamefont{Rado}} \bibnamefont{and}
  \bibinfo{editor}{\bibfnamefont{H.}~\bibnamefont{Suhl}}
  (\bibinfo{publisher}{Academic Press}, \bibinfo{address}{New York, London},
  \bibinfo{year}{1963}), vol. \bibinfo{volume}{III}.

\bibitem[{\citenamefont{Brown}(1963)}]{brown:1963PR}
\bibinfo{author}{\bibfnamefont{W.~F.} \bibnamefont{Brown}, \bibfnamefont{Jr.}},
  \bibinfo{journal}{Phys. Rev.} \textbf{\bibinfo{volume}{130}},
  \bibinfo{pages}{1677} (\bibinfo{year}{1963}).

\bibitem[{\citenamefont{Brown}(1959)}]{brown:1959JAP}
\bibinfo{author}{\bibfnamefont{W.~F.} \bibnamefont{Brown}, \bibfnamefont{Jr.}},
  \bibinfo{journal}{J. Appl. Phys.} \textbf{\bibinfo{volume}{30}},
  \bibinfo{pages}{S130} (\bibinfo{year}{1959}).

\bibitem[{\citenamefont{Bean and Livingston}(1959)}]{bean:1959JAP}
\bibinfo{author}{\bibfnamefont{C.~P.} \bibnamefont{Bean}} \bibnamefont{and}
  \bibinfo{author}{\bibfnamefont{J.~D.} \bibnamefont{Livingston}},
  \bibinfo{journal}{J. Appl. Phys.} \textbf{\bibinfo{volume}{30}},
  \bibinfo{pages}{S120} (\bibinfo{year}{1959}).

\bibitem[{\citenamefont{Hesse et~al.}(2000)\citenamefont{Hesse, Bremers, Hupe,
  Veith, Fritscher, and Valtchev}}]{hesse:2000JMMM}
\bibinfo{author}{\bibfnamefont{J.}~\bibnamefont{Hesse}},
  \bibinfo{author}{\bibfnamefont{H.}~\bibnamefont{Bremers}},
  \bibinfo{author}{\bibfnamefont{O.}~\bibnamefont{Hupe}},
  \bibinfo{author}{\bibfnamefont{M.}~\bibnamefont{Veith}},
  \bibinfo{author}{\bibfnamefont{E.~W.} \bibnamefont{Fritscher}},
  \bibnamefont{and} \bibinfo{author}{\bibfnamefont{K.}~\bibnamefont{Valtchev}},
  \bibinfo{journal}{J. Magn. Magn. Mater.} \textbf{\bibinfo{volume}{212}},
  \bibinfo{pages}{153} (\bibinfo{year}{2000}).

\end{thebibliography}

\end{document}